\documentclass[english,pra,aps,showpacs,twocolumn]{revtex4-1}

\def\be{\begin{equation}}
\def\ee{\end{equation}}
\usepackage{graphicx}
\usepackage{bm}
\usepackage[T1]{fontenc}
\usepackage[latin9]{inputenc}
\usepackage{amsmath}
\usepackage{amssymb}
\usepackage{float}
\usepackage[bookmarks=true,
   colorlinks=true,
   linkcolor=blue,
   urlcolor=blue,
   citecolor=blue,
   bookmarks=true,
   hyperindex=true
]{hyperref}

\begin{document}
\title{Expansion dynamics of a spherical Bose-Einstein condensate}

\author{Ruizong Li$^{1,4}$, Tianyou Gao$^{1}$, Dongfang Zhang$^{1}$, Shi-Guo Peng$^{1}$, Lingran Kong$^{1,4}$, Xing Shen$^{1,4}$, Wuming Liu$^{2}$}

\author{Kaijun Jiang$^{1,3}$}
\email{kjjiang@wipm.ac.cn}

\affiliation{$^{1}$State Key Laboratory of Magnetic Resonance and Atomic and Molecular Physics, Wuhan Institute of Physics and Mathematics, Chinese Academy of Sciences, Wuhan 430071, China}
\affiliation{$^{2}$Beijing National Laboratory for Condensed Matter Physics, Institute of Physics, Chinese Academy of Sciences, Beijing 100190, China}
\affiliation{$^{3}$Center for Cold Atom Physics, Chinese Academy of Sciences, Wuhan 430071, China}
\affiliation{$^{4}$School of Physics, University of Chinese Academy of Sciences, Beijing 100049, China}

\date{\today}

\begin{abstract}
We experimentally and theoretically observe the expansion behaviors of a spherical Bose-Einstein condensate. A rubidium condensate is produced in an isotropic optical dipole trap with an asphericity of 0.037. We measure the variation of the condensate size during the expansion process. The free expansion of the condensate is isotropic, which is different from that of the condensate usually produced in the anisotropic trap. The expansion in the short time is speeding and then after a long time the expansion velocity asymptotically approaches a constant value. We derive an analytic solution of the expansion behavior based on the spherical symmetry, allowing a quantitative comparison with the experimental measurement. The interaction energy of the condensate is gradually converted into the kinetic energy at the beginning of the expansion and the kinetic energy dominates after a long-time expansion. We obtain the interaction energy of the condensate in the trap by probing the expansion velocity, which is consistent with the theoretical prediction.
\end{abstract}

\maketitle

\section{Introduction}
In the ultracold Bose-Einstein condensate (BEC), the many-body interaction modifies the system behaviors deviate from the ideal gas \cite{Stringari1999RMP}, which provides a platform to explore exotic quantum effects such as low-energy excitations \cite{Wieman1996PRL, Ketterle1996PRL, Davison2005RMPexcitation}, phase transitions in optical lattices \cite{Bloch2008RMPlattice, Nori2014RMPsimulation, Eckardt2017RMPlattice}, matter-wave interferometry \cite{Phillips2000PRLphase, Schmiedmayer2009RMPinterferometry}, artificial gauge potential \cite{Spielman2011NATUREsoc, Dalibard2011RMPgauge}, low-dimension physics \cite{Weiss2006NATUREcradle, Giamarchi2011RMPonedimension}, and many others. Due to the small \textit{in-situ} size of the condensate in the trap, the cold atomic sample is usually probed after certain free expansion time \cite{Ketterle1999Arxiv}. Many-body interaction plays an important role to determine the expansion dynamics. After the condensate being released from the trap, the interaction energy is converted to the kinetic energy and the initial acceleration after switch-off of the trap is determined by the gradient of the interaction energy\cite{Stringari1996PRA, Baym1996PRL, Castin1996PRL}. The expansion behaviors of different directions are dependent on the configuration of the external trap. Previously the condensate is mostly produced in an anistotropic trap (i.e., $\epsilon\neq1$, where $\epsilon$ is the aspect ratio between the axial and radial frequencies) due to the technical challenge, which leads to an increased degree of complexity in the study of the expansion behavior. In this case the expansion behaviors couldn't be analytically solved without approximation because different coupled second-order differential equations are relevant \cite{Stringari1999RMP, Castin1996PRL, Cooper1997PRL, Pitaevskii1997PLAexpansion}. Obtaining the quantum system with an analytic solution allows a more lucid description of the condensate dynamics and an immediate comparison between experiment and theory. Spherical Bose condensate in an isotropic trap (i.e., $\epsilon=1$) is a special case that the expansion behavior can be analytically solved. Here we only need to solve one differential equation due to the spherical symmetry. Nevertheless, the experimental study of this expansion is still lacking.

The spherical condensate has unique features of the many-body interaction. The excitation spectrum of the condensate is simplified by degeneracy and becomes theoretically tractable \cite{Stringari1996PRL, Stringari1999RMP, Pitaevskii1999PRA}. The reduced availability of states in a spherical trap, caused by degeneracy, has a major effect on the Landau damping rate. Quantitative calculations of this process currently have been carried out \cite{Pitaevskii1997PLA, Pitaevskii1999PRA, Burnett2000PRLsphericalexcitation, Cornell2016PRA}. Accurately measuring the collective mode of the spherical condensate in the finite-temperature regime can extract subtle many-body effects like thermal and quantum fluctuation \cite{Hu2004PRAthermal, Griffin2001PRAthermal, Zaremba2002PRAthermal, Giorgini2000PRAthermal}.

Previously Hodby et al were able to modify the aspect ratio ($\epsilon=2.83 \sim 1.6$) in a magnetic trap while keeping the confinement tight \cite{Foot2000JPBtrap}. However they do not report achieving a fully isotropic trap. Lobser and his colleagues realized an isotropic magnetic trap with the aid of the gravity force \cite{Cornell2015NaturePhy, Cornell2016PRA}. But the weak confinement (the trapping frequency $\omega \approx 2\pi\times9$ Hz) in their work is disadvantageous to obtain the pure condensate.

In this paper we produce a nearly spherical rubidium condensate in an optical dipole trap with an asphericity of 0.037. The large trapping frequency ($2\pi\times77.5$ Hz) is favorite to produce a pure condensate. Then we measure the condensate widths during the expansion process as well as the interaction energy of the condensate in the trap. The experimental results are consistent with the theoretical predictions based on the spherical symmetry. We explore the expansion dynamic process in which the the interaction energy is converted into the kinetic energy and becomes smaller versus the expansion time.

The paper is organized as follows. We first present the production of a spherical rubidium condensate in Sec. II. Then we introduce the expansion behaviors of the condensate in Sec. III. Subsequently the interaction energy of the condensate in the trap is obtained in Sec. IV. Finally, the conclusions are summarized in Sec. V.

\section{Production of a spherical Bose condensate}
The experimental configuration is composed of double magneto-optical traps (MOTs), which is similar to our previous works \cite{Jiang2016CPLbec, Jiang2018CPLplug}. $^{87}$Rb atoms are cooled and trapped in the first MOT and then transferred to the second MOT with a series of optical pushing pulses. In the second MOT the atom number is $8.5(9)\times10^8$ and the atom temperature is $320(40)$ $\mu$K. The atom temperature is reduced to 130(20) $\mu$K after a decompressed MOT process. Then the atoms are loaded into a magnetic trap by scanning the magnetic gradient to 336.0 G/cm with a period of 300 ms. Here the atom number is $2.0(7)\times10^8$ and the atom temperature is 210(25) $\mu$K. The atoms are cooled with the ratio frequency (RF) induced evaporation cooling to 15(3) $\mu K$ and subsequently transferred into a hybrid trap composed of magnetic and optical dipole fields \cite{Spielman2009PRA}. Finally we transfer cold atoms into an optical dipole trap by gradually decreasing the magnetic trap.

\begin{figure}
\centerline{\includegraphics[width=8.5cm]{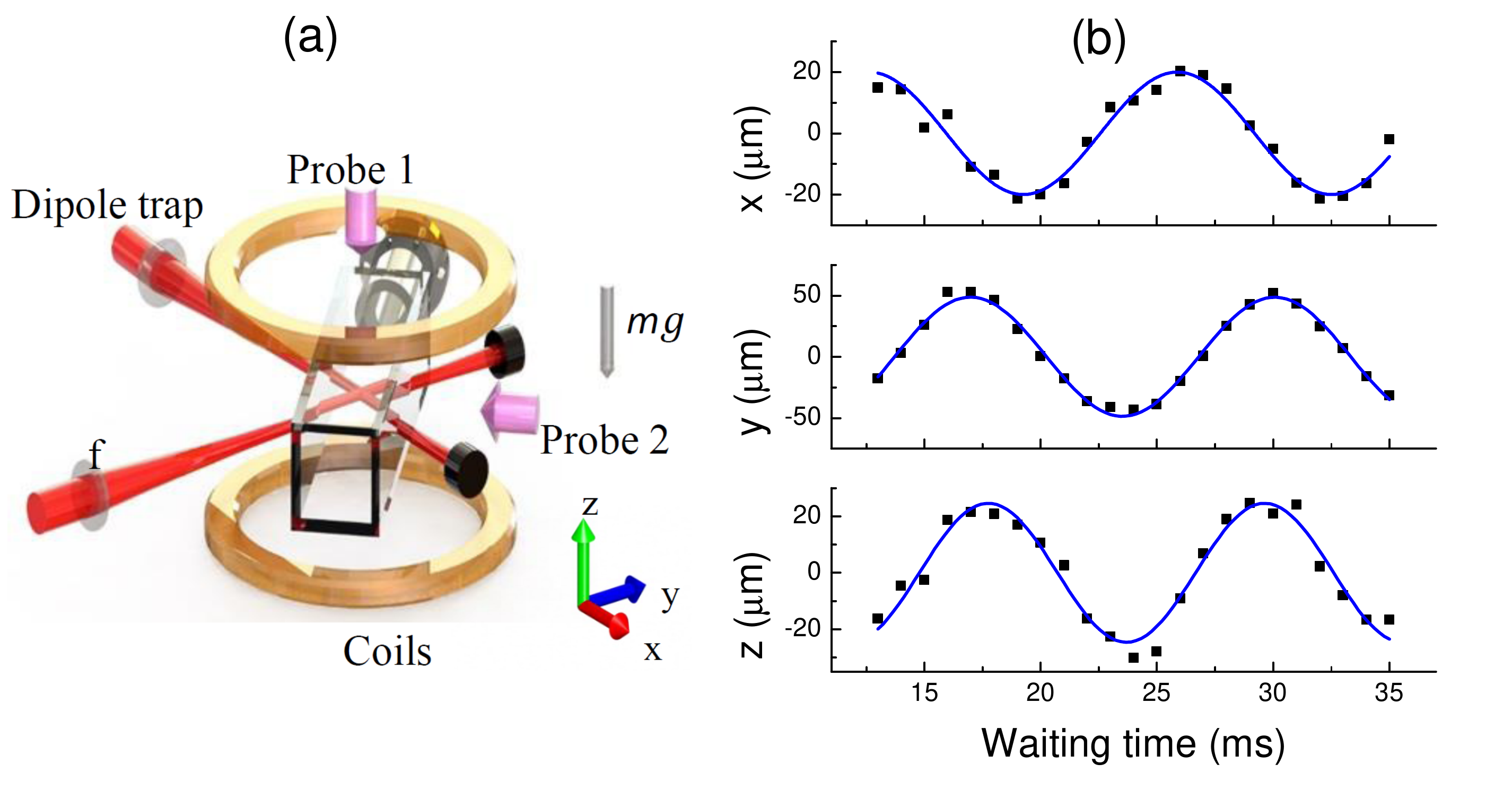}}
\caption{(color online) (a) Experimental setup. The optical dipole trap is composed of two focused red-detuned laser beams in $x$ and $y$ directions. The gravity is in the $-z$ direction. Ultracold atoms are simultaneously probed in the vertical and horizontal directions. (b) Measuring the trapping frequencies by probing the oscillations of the centers of mass along three directions, respectively. Each experimental data is the average of three measurements. The solid line is the fitting with a sine wavefunction.}  \label{Fig1}
\end{figure}

We produce a spherical $^{87}$Rb BEC in an optical dipole trap in which the trapping frequencies along $x,y,z-$directions are the same. As shown in Fig.~\ref{Fig1}(a), the optical dipole trap is composed of two far red-detuned laser beams with the wavelength $\lambda=1064$ nm. The Rayleigh length, $z_{R}=\pi w_{0}^{2}/\lambda$, is much larger than the beam waist $w_{0}$. For a single laser beam, the trapping frequency in the radial direction is about 200 times larger than that in the axial direction. So the trapping effect along the propagation direction can be neglected. To produce a fully isotropic trap, the gravity force should be included \cite{Cornell2015NaturePhy, Foot2000JPBtrap}. The trapping potential, which is composed of the optical dipole trap and the gravity, is given by

\begin{equation} \label{eq:trap}
\begin{split}
U \left(x,y,z\right) =& -U_{1}\exp\left(-\frac{2x^{2}}{w_{1x}^{2}}-\frac{2z^{2}}{w_{1z}^{2}}\right)\\
&-U_{2}\exp\left(-\frac{2y^{2}}{w_{2y}^{2}}-\frac{2z^{2}}{w_{2z}^{2}}\right)-mgz.
\end{split}
\end{equation}

\noindent $w_{1x}$ ($w_{2y}$) and $w_{1z}$ ($w_{2z}$) are the waists of the optical beam along the $y$ ($x$) direction, and $U_1$ and $U_2$ are the peak potential energies of the two beams, respectively. By expanding Eq. (\ref{eq:trap}) in the potential minimum $(0,0,z_{0})$ to the second order, forming a spherical BEC should satisfy the conditions

\begin{eqnarray}
U_{1}=\frac{mgw_{1x}^{2}}{2a}\sqrt{\frac{w_{1x}^{2}/w_{1z}^{4}+w_{2y}^{2}/w_{2z}^{4}}{a-1}} \label{eq:trap1}\\
U_{2}=\frac{mgw_{2y}^{2}}{2a}\sqrt{\frac{w_{1x}^{2}/w_{1z}^{4}+w_{2y}^{2}/w_{2z}^{4}}{a-1}} \label{eq:trap2}
\end{eqnarray}

\noindent where $a=w_{1x}^{2}/w_{1z}^{2}+w_{2y}^{2}/w_{2z}^{2}$. In the experiment, we can accurately adjust the intensities of the two beams to simultaneously match Eq. (\ref{eq:trap1}) and Eq. (\ref{eq:trap2}).

By displacing atoms away from the equilibrium position for 2 ms, we measure the oscillation of the center of the mass (COM) of the atomic cloud in the trap along the $x$, $y$ and $z$ directions, respectively. The experimental results are shown in Fig.~\ref{Fig1}(b). Using a sine wavefunction $r_{i}=A_{i}\sin(\omega_{i}t+\phi_{i})$ ($i=x, y, z$ and $r_{i}\rightarrow i$) to fit the experimental data, we get the trapping frequencies: $\omega_{x}= 2\pi\times 76.7(14)$ Hz, $\omega_{y}= 2\pi\times 76.5(6)$ Hz, $\omega_{z}= 2\pi\times 79.4(12)$ Hz. The frequency uncertainties in the parenthesis are from the fitting process. So the mean trapping frequency $\bar{\omega}= \left(\omega_{x}+\omega_{y}+\omega_{z}\right)/3 = 2\pi\times 77.5$ Hz which is much larger than that in the reference \cite{Cornell2015NaturePhy}. The asphericity $A=(\omega_{max}-\omega_{min})/\bar{\omega}\approx 0.037$, where $\omega_{max}$, $\omega_{min}$ are the maximum and minimum trapping frequencies along three directions, respectively. The tight confinement here is favorite to produce a pure Bose condensate with negligible thermal gases and obtain experimental data with a high signal-to-noise ratio. We improve the position stability of the optical trap beam better than 3 $\mu$m to achieve a stable spherical BEC. The atoms stay in the spin state $|F, m_{F}\rangle = |1, -1\rangle$. The atom number is about $1.2\times10^5$. The BEC is well in the hydrodynamic limit with the adimensional parameter $N a_s/a_{ho}\approx 570\gg 1$~\cite{Stringari1996PRL}, where $a_s$ is the s-wave scattering length, $a_{ho}=\sqrt{\hbar/m \omega}$ is the harmonic oscillator length of the trap, and $N$ is the atom number.

\begin{figure}
\centerline{\includegraphics[width=7cm]{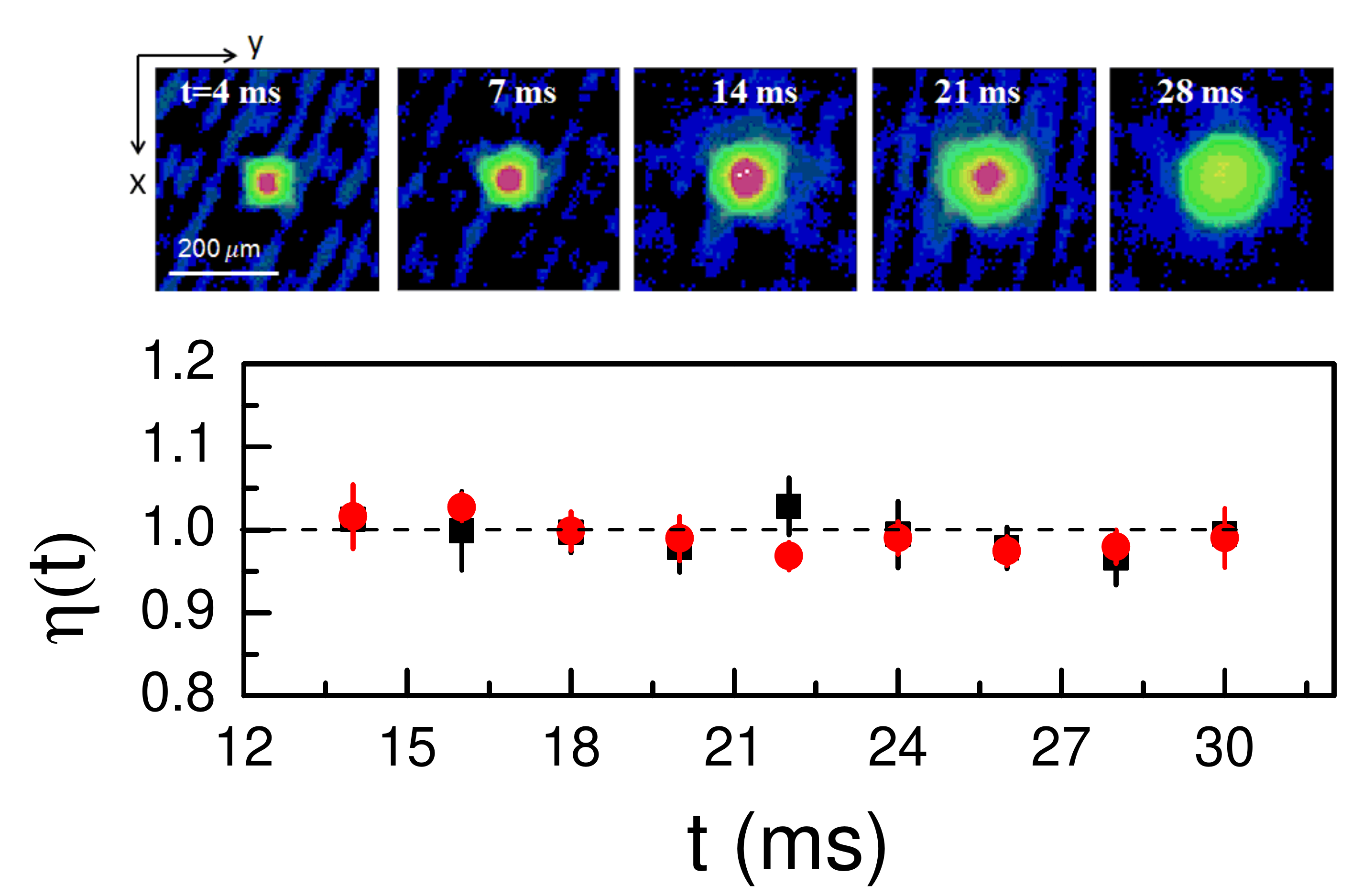}}
\caption{(Color online) Isotropic expansion of the condensate. The upper row shows exemplary images probed in the vertical direction for five expansion times. The lower row shows the aspect ratio $\eta(t)$ versus the expansion time $t$. The black squares (red circles) are for the images probed in the horizontal (vertical) direction. Each error bar indicates the uncertainty of three measurements. The dashed line denotes the value of unity. }
\label{Fig2}
\end{figure}

After suddenly switching off the optical trap, we measure the aspect ratio $\eta(t)$ of the condensate during the free expansion. The experimental results are shown in Fig.~\ref{Fig2}. The condensate width $R_{i}(t)$ during the expansion is obtained by fitting the optical density of the image with a Thomas-Fermi (TF) distribution along the corresponding direction. For the images probed in the horizontal direction as shown in Fig.~\ref{Fig1}(a), $\eta(t)=R_{\parallel}(t)/R_z(t)$ where $R_{\parallel}(t)$ and $R_z(t)$ are the TF radii in the horizontal and vertical directions, respectively. For the images probed in the vertical direction, $\eta(t)=R_{x}(t)/R_y(t)$ where $R_{x}(t)$ and $R_y(t)$ are the TF radii in the $x$ and $y$ directions, respectively. $\eta(t)$ remains unity during the free expansion, which is unique for a spherical BEC. For an anisotropic BEC, the expansion is anisotropic and the aspect ratio $\eta(t)$ approaches an asymptotic value dependent on the ratio of the trapping frequencies \cite{Rempe1998EPL, Rempe1998APB, Ketterle1996PRLrelease, Stringari1999RMP, Castin1996PRL}.

\section{Expansion behavior of the condensate}

\begin{figure}
\centerline{\includegraphics[width=8.5cm]{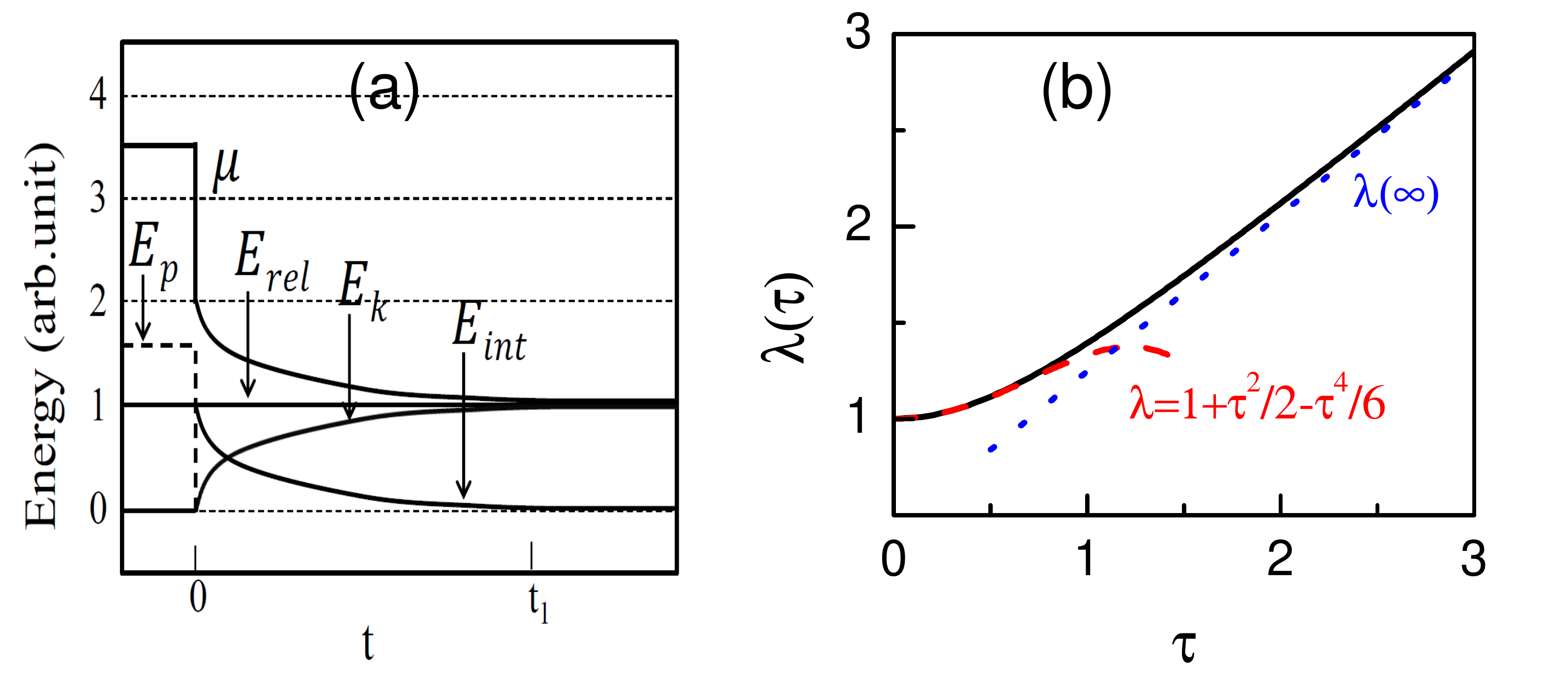}}
\caption{(Color online) (a) Schematics of energy components during the free expansion. The trapping potential is switched off at $t=0$. $E_{p}$ is the potential energy, $E_{rel}$ is the release energy, $E_{kin}$ is the kinetic energy, $E_{int}$ is the interaction energy and $\mu$ is the chemical potential. After a long-time expansion ($t>t_{1}$), interaction energy is completely converted to kinetic energy. (b) The scaling factor $\lambda(\tau)=R(\tau)/R(0)$ versus the scaling expansion time $\tau$. $R(0)$ is the TF radius of the BEC in the trap and $\tau = \omega t$. The black solid curve denotes the calculation with Eq.(\ref{eq:exp11}) for the whole expansion process. The red dashed curve indicates the calculation with Eq.(\ref{eq:exp2}) for the short-time expansion. The blue dotted curve is the calculation with Eq.(\ref{eq:exp3}) for the long-time expansion . }\label{Fig3}
\end{figure}

Many-body interaction plays an important role to determine the expansion dynamics of BEC. Fig.~\ref{Fig3}(a) briefly indicates variations of different energy components during the expansion process \cite{Stringari1999RMP, Stringari1996PRA, Cooper1997PRL}. The chemical potential $\mu=E_{kin}+E_{p}+2E_{int}$ is composed of kinetic energy $E_{kin}$, potential energy $E_{p}$ and interaction energy $E_{int}$. In the trap, $E_{kin}$ is negligibly small and $E_{p}=1.5E_{int}$ according to the Virial relation $2E_{kin}-2E_{p}+3E_{int}=0$. After BEC being released from the trap, $E_{p}$ is switched off and $E_{int}$ starts to be converted into $E_{kin}$ gradually, which makes the release energy $E_{rel}=E_{kin}+E_{int}$ keep constant during the expansion. After a long-time expansion, the interaction energy is completely converted to the kinetic energy. This provides a efficient way to measure the interaction energy of BEC in the trap by probing the expansion velocity in the long-time expansion, which will be followed in section IV.

During the expansion, the atomic cloud experiences just a free dilatation. Three scaling factors, i.e., $\lambda_{i}\left(t\right)\equiv r_{i}\left(t\right)/r_{i}\left(0\right)$($i=x,y,z$), may be introduced as in references \cite{Stringari1999RMP, Castin1996PRL, Rempe1998APB, Rempe1998EPL}, which describe the trajectory of any infinitesimally small fraction at the position ${\bf{r}}\left(t\right)$ of the moving cloud. For an axially symmetric BEC mostly produced previously (i.e., $\lambda_{x}(t)=\lambda_{y}(t)\neq \lambda_{z}(t)$), it is required to solve two coupled second-order differential equations to get the evolution of the scaling factors \cite{Stringari1999RMP, Castin1996PRL, Cooper1997PRL, Pitaevskii1997PLAexpansion}. In this case analytic solution is generally absent. Here for the spherical BEC with $\lambda(t)=\lambda_{x}(t)=\lambda_{y}(t)= \lambda_{z}(t)$, the previous two coupled differential equations simply merge into one,
\begin{equation}
\frac{\partial^{2}\lambda}{\partial \tau^{2}}=\lambda^{-4},
\end{equation}
whose solution can be obtained analytically,
\begin{equation}
\tau = -\sqrt{\frac{3}{2}}\cdot\frac{\sqrt{\pi}\Gamma(2/3)}{\Gamma(1/6)} + \sqrt{\frac{3}{2}}\lambda \cdot\, _{2}F_{1}\left(-\frac{1}{3},\frac{1}{2},\frac{2}{3},\frac{1}{\lambda^{3}}\right), \label{eq:exp11}
\end{equation}
where $\tau=\omega t$, $\Gamma(\cdot)$ is the Gamma function, and $_{2}F_{1}(a,b,c,z)$ is the hypergeometric function.

According to Eq.(\ref{eq:exp11}), we can easily obtain the asymptotic behavior of the scaling factor $\lambda\left(\tau\right)$ for a short- or long-time expansion,
\begin{eqnarray}
\lambda (\tau) \approx 1+\tau^2/2- \tau^4/6, (\tau \rightarrow 0)  \label{eq:exp2}\\
\lambda (\tau) \approx \sqrt{2/3} \tau+\frac{\sqrt{\pi}\Gamma(1/6)}{\Gamma(2/3)}. (\tau \rightarrow \infty)  \label{eq:exp3}
\end{eqnarray}

The scaling factor of the condensate during the expansion process is shown in Fig.~\ref{Fig3}(b). For the short-time expansion ($\tau\rightarrow 0$), ${\partial \lambda}/{\partial \tau} \approx \tau$ and ${\partial^2 \lambda}/{\partial \tau^2}=1-2\tau^2$. This means that after the trap being switched off, the expansion is speeding but the acceleration decreases versus the expansion time. These behaviors can be explained that the interaction energy is gradually converted into the kinetic energy and decreases versus the expansion time \cite{Ketterle1996PRLrelease, Cooper1997PRL, Stringari1999RMP}. For the long-time expansion ($\tau \rightarrow \infty$), ${\partial \lambda}/{\partial \tau} = \sqrt{2/3}$. It turns out that the interaction energy has been completely converted to the kinetic energy and the expansion velocity finally reaches a constant value. The quantitative calculations in Fig.~\ref{Fig3}(b) can clearly confirm the expansion dynamics as shown in Fig.~\ref{Fig3}(a).

\begin{figure}
\centerline{\includegraphics[width=7cm]{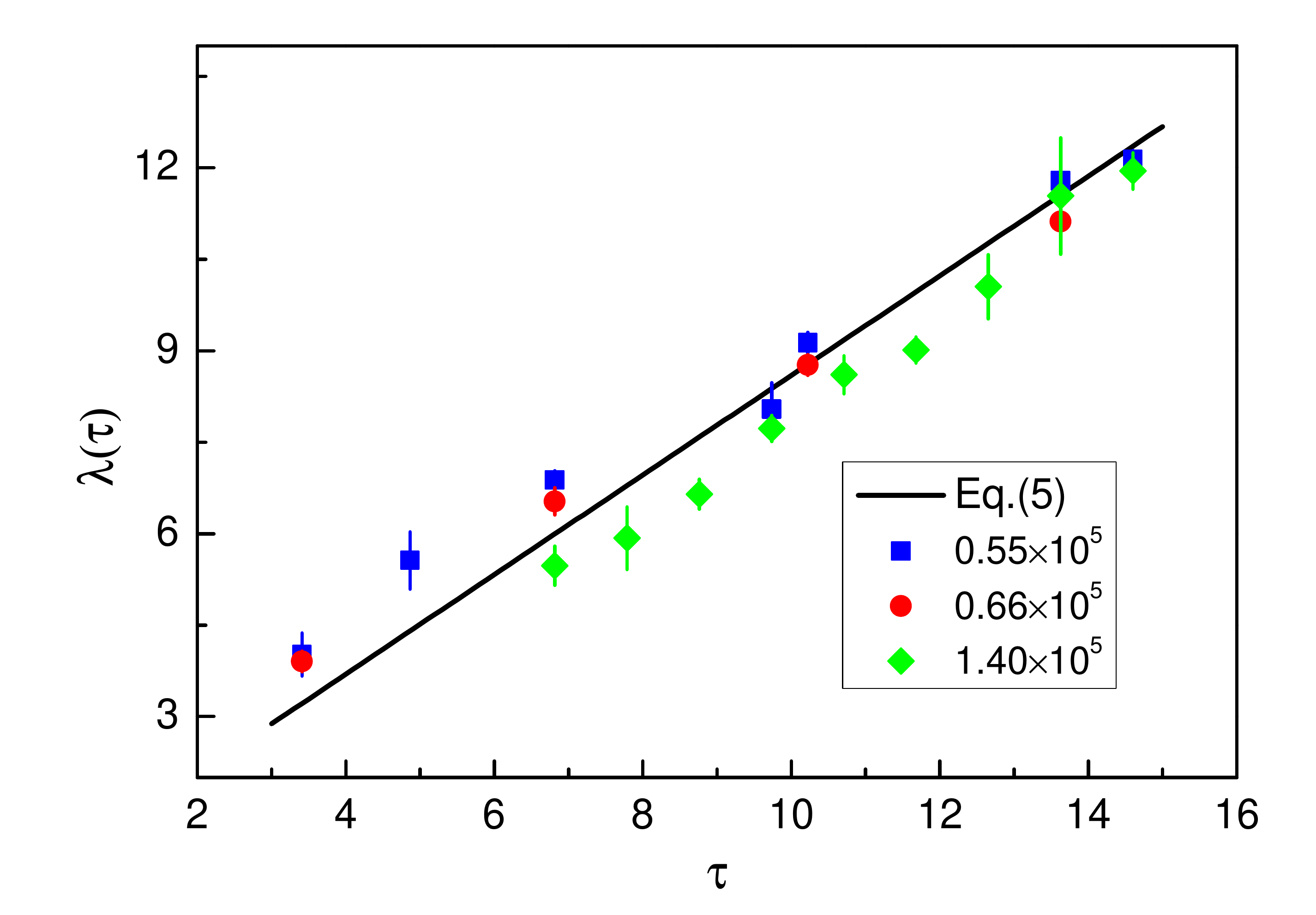}}
\caption{(Color online) Scaling factor $\lambda(\tau)$  versus the scaling expansion time $\tau$. The black solid curve is the calculation with Eq.(\ref{eq:exp11}). Blue squares, red circles and green diamonds denote the measurements with atom numbers of $0.55 \times 10^5$, $0.66 \times 10^5$ and $1.40 \times 10^5$. Each error bar is the uncertainty of three measurements.}\label{Fig4}
\end{figure}

In Fig. \ref{Fig4} we measure the scaling factor during the expansion, which is extracted by measuring the size of the atomic cloud after some expansion time. Due to the limited resolution of the imaging system ($\Delta r \approx 7.6$ $\mu$m) \cite{Jiang2018CPLplug}, we only show the experimental data for expansion times larger than 7 ms. Here the size of the cloud is defined as $R(\tau)=\left[R_{x}(\tau)+R_{y}(\tau)+R_{z}(\tau)\right]/3$, where $R_{i}(\tau)$ ($i=x,y,z$) is the TF radius along the principle axis, and $\lambda(\tau)=R(\tau)/R(0)$. Under the TF approximation, $R(0)$ is calculated from the atom number and the trapping frequency, and $R(\tau)$ is measured in the experiment. Experimental results at three atomic numbers are consistent with the theoretical prediction of Eq. (\ref{eq:exp11}).

\section{Interaction energy of the condensate}
During the expansion, the interaction energy is gradually converted into the kinetic energy $E_{kin}$. After atoms being releasing from the trap, $E_{kin}$ can be written in the following integral
\begin{equation}
E_{kin}=\int\frac{1}{2}m\left[\frac{d{\bf r}\left(t\right)}{dt}\right]^{2}n\left[{\bf r}\left(t\right)\right]\mathcal{D}\left[{\bf r}\left(t\right)\right],\label{eq:4.1}
\end{equation}
 where $n\left[{\bf r}\left(t\right)\right]$ is the density at the
position ${\bf r}\left(t\right)$, and $m$ is the atomic mass. Using
${\bf r}\left(t\right)=\lambda\left(t\right){\bf r}\left(0\right)$,
Eq.(\ref{eq:4.1}) becomes
\begin{equation}
E_{kin}=\frac{1}{2}m\left[\frac{d\lambda\left(t\right)/dt}{\lambda\left(t\right)}\right]^{2}\int{\bf r}\left(t\right)^{2}n\left[{\bf r}\left(t\right)\right]\mathcal{D}\left[{\bf r}\left(t\right)\right].\label{eq:4.2}
\end{equation}
It was shown in \cite{Castin1996PRL} that the density $n\left({\bf r}\right)$ still satisfies the
generalized TF distribution for a time-dependent problem, which takes the form of
\begin{equation}
n\left[{\bf r}\left(t\right)\right]=\frac{15}{8\pi R\left(t\right)^{3}}\left[1-\frac{r\left(t\right)^{2}}{R\left(t\right)^{2}}\right],\label{eq:4.3}
\end{equation}
for a spherical atomic cloud, where $R\left(t\right)$ is the generalized TF radius at
the time $t$, $n\left[{\bf r}\left(t\right)\right]$ has been normalized
to unity, i.e., $\int_{\mathcal{S}}n\left[{\bf r}\left(t\right)\right]\mathcal{D}\left[{\bf r}\left(t\right)\right]=1$,
and $\mathcal{S}$ is a spherical domain with the radius $R\left(t\right)$.
Substituting Eq.(\ref{eq:4.3}) into (\ref{eq:4.2}), we easily obtain
\begin{equation}
E_{kin}=\frac{1}{2}m\cdot\frac{3}{7}\left[\frac{dR\left(t\right)}{dt}\right]^{2}.\label{eq:4.4}
\end{equation}

As shown in Fig. \ref{Fig3}, the interaction energy is completely converted to the kinetic energy after a long-time expansion. So the interaction energy $E_{int}$ at $t\rightarrow 0$ is roughly equivalent to the kinetic energy $E_{kin}$ at $t\rightarrow \infty$.

We can check the validity of Eq.(\ref{eq:4.4}). From Eq.(\ref{eq:exp3}) the size of the cloud $R\left(t\right)$ as $t\rightarrow\infty$ should behave as
\begin{equation}
R\left(t\right)\approx\sqrt{\frac{2}{3}}\omega tR\left(0\right),\label{eq:2.7}
\end{equation}
where $R\left(0\right)$ is the size at $t=0$. Then the interaction energy of the condensate in the trap becomes
\begin{equation}
E_{int}=\frac{1}{2}m\cdot\frac{3}{7}\cdot\frac{2}{3}\omega^{2}R\left(0\right)^{2}=\frac{2}{7}\cdot\frac{1}{2}m\omega^{2}R\left(0\right)^{2}=\frac{2}{7}\mu,\label{eq:2.8}
\end{equation}
and $\mu=m\omega^{2}R\left(0\right)^{2}/2$ is the chemical potential in the trap. Eq.(\ref{eq:2.8}) is consistent with the well-known result $E_{int}=2\mu/7$ \cite{Stringari1999RMP}.

\begin{figure}
\centerline{\includegraphics[width=7cm]{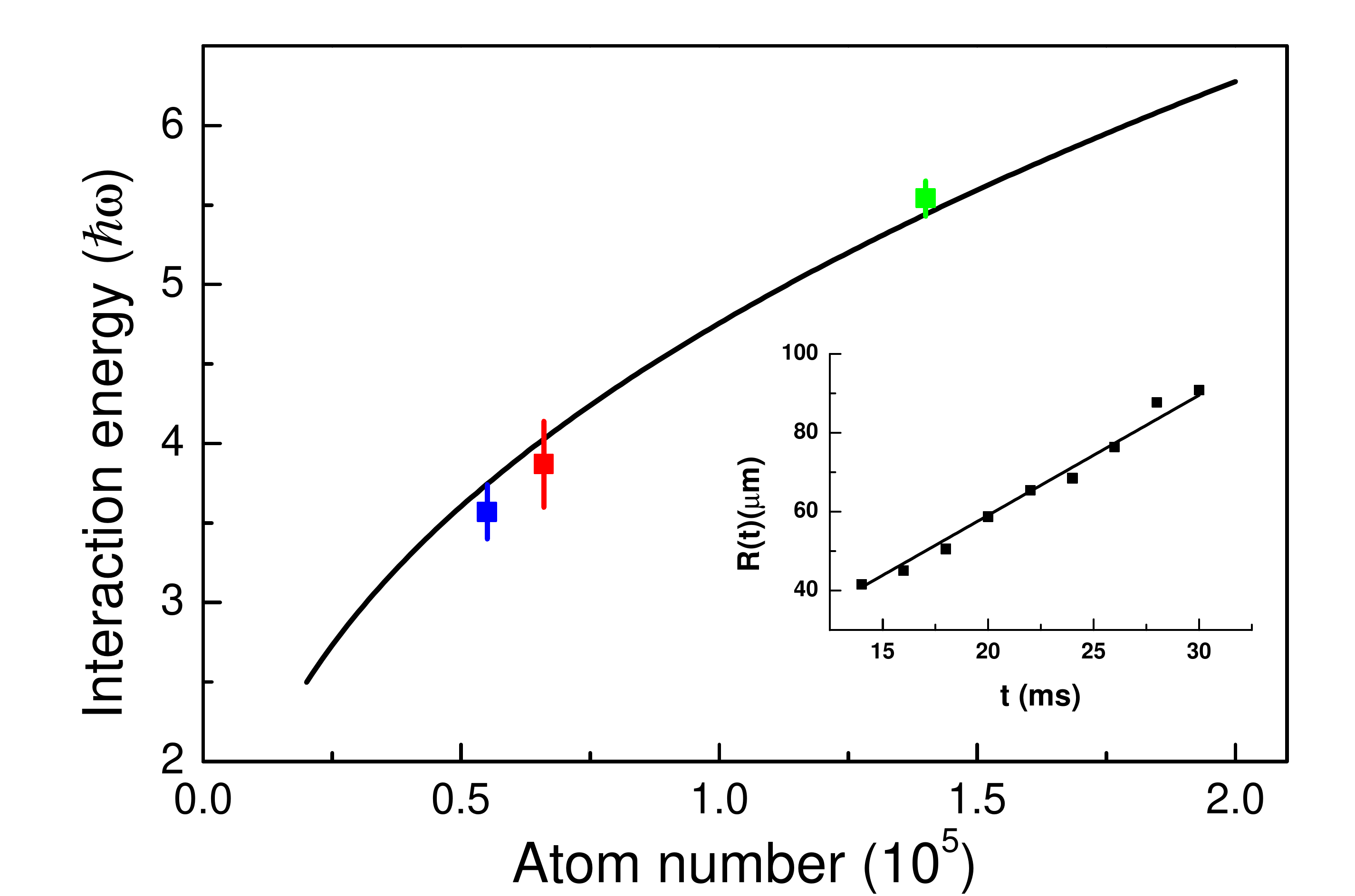}}
\caption{(Color online) Interaction energy of the condensate in the trap versus the atom number. The solid curve is the theoretical calculation with Eq. (\ref{eq:2.9}).  The measurements are for atom numbers of $0.55 \times 10^5$, $0.66 \times 10^5$ and $1.40 \times 10^5$. The error bar is the uncertainty in fitting the expansion velocity. The insert shows an example for linearly fitting the measured TF radii of the condensate with the atom number of $1.40 \times 10^5$. The expansion times are larger than 14 ms. }\label{Fig5}
\end{figure}

In the experiment we extract the expansion velocity for the long-time expansion by linearly fitting the TF radii of the condensate. One example of this fitting process is shown in the inset of Fig. \ref{Fig5}. The expansion times are larger than $t>14$ ms (i.e., $\tau>6.8$), which ensures that the expansion velocity has approached the constant value (see Fig.~\ref{Fig3}(b)). Then the interaction energy of the condensate in the trap can be calculated with Eq. (\ref{eq:4.4}).

On the other hand, the chemical potential of the condensate in the trap can be calculated with \cite{Stringari1996PRA, Baym1996PRL, Stringari1999RMP, Castin1996PRL}

\begin{equation}
\mu=\frac{\hbar \omega}{2}(\frac{15Na_{s}}{a_{ho}})^{2/5}.\label{eq:2.9}
\end{equation}
The interaction energy of the condensate in the trap versus the atom number is ploted  in Fig. \ref{Fig5}. The experimental measurements are consistent with the theoretical prediction with Eq. (\ref{eq:2.9}).

\section{Conclusions}
In conclusion, we experimentally observe the expansion behaviors of a spherical Bose condensate. A spherical rubidium condensate is produced in an optical dipole trap and the characteristic isotropic expansion is observed in the experiment. The condensate widths during the expansion process as well as the interaction energy of the condensate in the trap are measured. The analytic solution of the expansion behavior of the condensate is derived, as a quantitative comparison with the experimental measurements. We find that the expansion in the short time is speeding and then after a long time the expansion velocity is constant. The intrinsic mechanics of this behavior is that the interaction energy is converted into the kinetic energy at the beginning of the expansion and the kinetic energy dominates after a long-time expansion.

This work has already been accessible to study the frequency shift and Landau damping of low-energy excitations in a spherical condensate, which allows the immediate comparison between the experiment and theory due to the simplified excitation spectrum by degeneracy \cite{Stringari1996PRL, Stringari1999RMP, Pitaevskii1999PRA, Pitaevskii1997PLA, Pitaevskii1999PRA, Burnett2000PRLsphericalexcitation, Cornell2016PRA, Hu2004PRAthermal, Griffin2001PRAthermal, Zaremba2002PRAthermal, Giorgini2000PRAthermal}. Compared to a magnetic trap, the spherical trap composed of the optical field is advantageous to study the non-equilibrium dynamics with the spherical symmetry, where fast modulation of the confinement strength is required.

\section*{Acknowledgments}
This work has been supported by the NKRDP (National Key Research and Development Program) under Grant No. 2016YFA0301503, NSFC (Grant No. 11674358, 11434015, 11474315) and CAS under Grant No. YJKYYQ20170025.

\end{document}